\documentclass[aps,superscriptaddress,showpacs,
               reprint]{revtex4-1}

\usepackage[utf8]{inputenc}
\usepackage{amsmath,amsfonts,amssymb}
\usepackage{pifont}
\usepackage{rotating}
\usepackage{graphicx}    
\usepackage{epstopdf}
\usepackage{color}
\usepackage[caption=false]{subfig}
\usepackage{soul}

\usepackage{hyperref}
\definecolor{dark-red}{rgb}{0.9,0.15,0.15}
\definecolor{dark-blue}{rgb}{0.15,0.15,0.4}
\definecolor{medium-blue}{rgb}{0,0,0.5}
\hypersetup{
    colorlinks, linkcolor={black},
    citecolor={dark-blue}, urlcolor={medium-blue}
}

\begin{document}

\title{Investigation of the phase separation property in La$_{0.2}$Pr$_{0.4}$Ca$_{0.4}$MnO$_{3}$ manganite}
\author{Deepak Kumar} 
\affiliation{Low Temperature Laboratory, Department of Physics, Indian Institute of Technology Bombay, Mumbai 400076, India}
\author{Ashwin A. Tulapurkar}
\affiliation{Department of Electrical Engineering IIT Bombay, Powai Mumbai 400076, India}

\author{C.V.Tomy} 
\email{tomy@phy.iitb.ac.in}
\email{kumar1990deepakk@gmail.com}
\affiliation{Low Temperature Laboratory, Department of Physics, Indian Institute of Technology Bombay, Mumbai 400076, India}

\begin{abstract}
We report a comprehensive investigation of La$_{0.2}$Pr$_{0.4}$Ca$_{0.4}$MnO$_{3}$ to clarify the micrometre scale phase separation phenomenon in the mixed valent manganite (La,Pr,Ca)MnO$_{3}$. The compound shows multiple magnetic transitions, in which the charge-ordered state is converted into a ferromagnetic state in steps with the application of a magnetic field. The ac susceptibility measurements show that the glassy transition at low temperatures does not depend on the frequency, thus indicating the absence of any spin glass behaviour. Magnetization as well as heat capacity measurements indicate that this low temperature transition is magnetic field dependent. The field dependent resistivity at 2\,K shows a sharp drop indicating that the sample behaviour changes from a high resistive state to a low resistive state, corroborating the conversion of charge-ordered insulating (COI) phase to a ferromagnetic metallic (FMM) phase. Our results point towards the existence of phase separation, rigidity of the low temperature glassy-like phase as well as the conversion of COI phase to FMM phase by the application of magnetic fields.
   
\end{abstract}

\date{\today}

\maketitle

\section{Introduction}

Mixed-valent manganites of the type $R_{1-x}A_{x}$MnO$_{3}$ ($R$ is a trivalent rare earth metal and $A$ is a divalent alkaline earth metal) illustrate a variety of physical phenomena such as metal to insulator transition (MIT), charge/orbital ordering (CO/OO), etc., in addition to the well-known colossal magnetoresistance (CMR) property \cite{Coey,TOKURA,Yakubovskii,Srikanth,Himanshu,Deepak,Tulapurkar} due to the coexistence of competing magnetic and electronic states. In addition, the manganites show the interesting phase separation (PS) phenomenon, which has been studied extensively, especially in La$ _{1-x-y} $Pr$ _{y} $Ca$ _{x} $MnO$ _{3} $ (LPCMO) system which is considered to be a classical prototype \cite{UeharaNature,Sharma,Mori,Macia}. For Ca doping  near $x=3/8$, the sample at low temperatures is a mixture of a ferromagnetic metal (FMM) (La$ _{5/8} $Ca$ _{3/8} $MnO$ _{3} $; $y=0$) and a charge ordered antiferromagnetic insulator (CO-AFI) (Pr$ _{5/8} $Ca$ _{3/8} $MnO$ _{3} $; $y=5/8$) \cite{Srikanth}. One of the key issues which was probed in detail was the glassy nature of the PS state (strain liquid to strain glass) due to the frustrated interactions in the LPCMO system. Even after extensive research corroborating the PS phenomenon in the LPCMO system \cite{UeharaNature,Jeen,Diaz,Kiryukhin}, the physical origin of the micrometre scale phase separation \cite{Bhalla}, with multiple coexisting phases, is not well understood due to the energetically comparable COI and FMM phases. In this paper we report  the heat capacity and ac susceptibility of the polycrystalline La$_{0.2}$Pr$_{0.4}$Ca$_{0.4}$MnO$_{3}$ sample in addition to the magnetization and transport measurements as a function of magnetic field and temperature. This chosen Pr stoichiometry ($y=0.4$) is shown to have disorder induced glassy (frozen) state in the system and a pronounced PS characteristic \cite{Ghivelder} at low temperatures. We have prepared the sample with $x=0.4$ in order to make sure that we are close to the boundary where the low temperature canted AFM state exists, without altering any other characteristics.

\section{Experimental details}

The La$_{0.2}$Pr$_{0.4}$Ca$_{0.4}$MnO$_{3}$ (LPCMO) sample in polycrystalline form was prepared by the standard solid  state reaction  method. The oxides La$_{2}$O$_{3}$, Pr$_{6}$O$_{11}$, CaCO$_{3}$ and MnO$_{2}$ used for preparing the sample, were preheated for removing the moisture before weighing. After preheating, the oxides were mixed in a proper stoichiometric ratio and ground in a mortar and a pestle. The mixture was then calcined at 1200$^{\circ}$C for 24 hrs. The calcined powder was pelletized and sintered at 1400$^{\circ}$C for 24 hours. The structure was characterized by the Rietveld analysis of the x-ray powder diffraction data with the help of Full-Prof software. The magnetization measurements were performed using a SQUID-VSM (Quantum Design Inc., USA) in the temperature range 2--300\,K with applied magnetic fields up to $\pm$70\,kOe. The resistivity was measured using a standard four-probe method in the temperature range 2--300\,K at different fields using a physical property measurement system (PPMS, Quantum Design, USA).

\section{Results and Discussion}
\begin{figure}[hbtp]
	\centering
	\includegraphics[width=\linewidth]{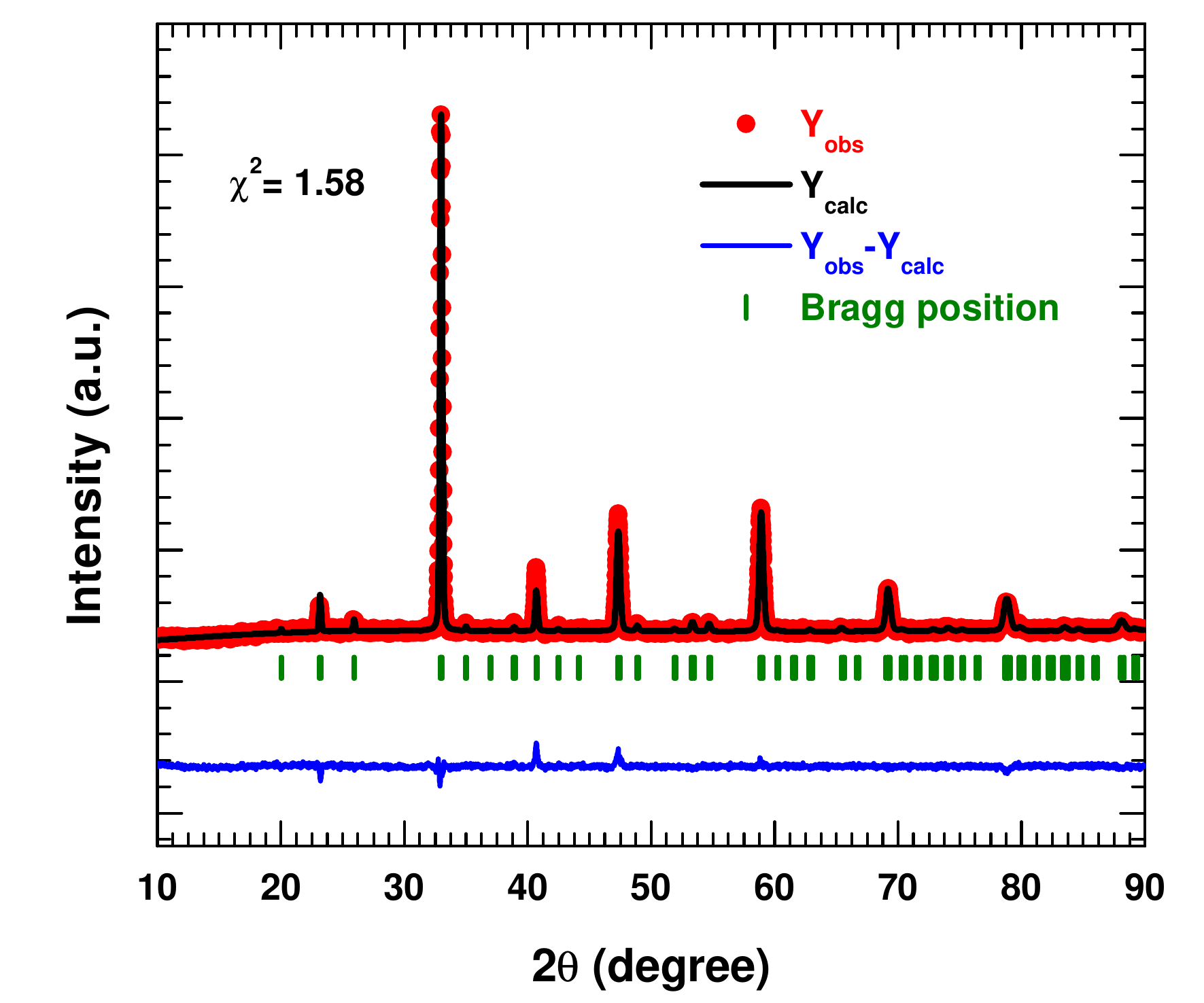}
	\caption{Rietveld-refined XRD pattern of La$_{0.2}$Pr$_{0.4}$Ca$_{0.4}$MnO$_{3}$ at room temperature.}
	\label{fig:XRD}
\end{figure}

Figure~\ref{fig:XRD} shows the Rietveld-refined (goodness of fit  $ \mathrm{\chi^{2}=1.58} $) XRD pattern of La$_{0.2}$Pr$_{0.4}$Ca$_{0.4}$MnO$_{3}$ at room temperature, which clearly shows that the sample LPCMO forms in a single phase. The structure of the sample is orthorhombic with a space group belonging to $ Pnma $. The lattice parameters obtained from the fit are $ a = 5.439$\,\AA, $b=7.667$\,\AA, $c=5.431$\,\AA, which match well with the reported values \cite{Collado}.

\begin{figure}[hbtp]
 	\centering
 	\includegraphics[width=\linewidth]{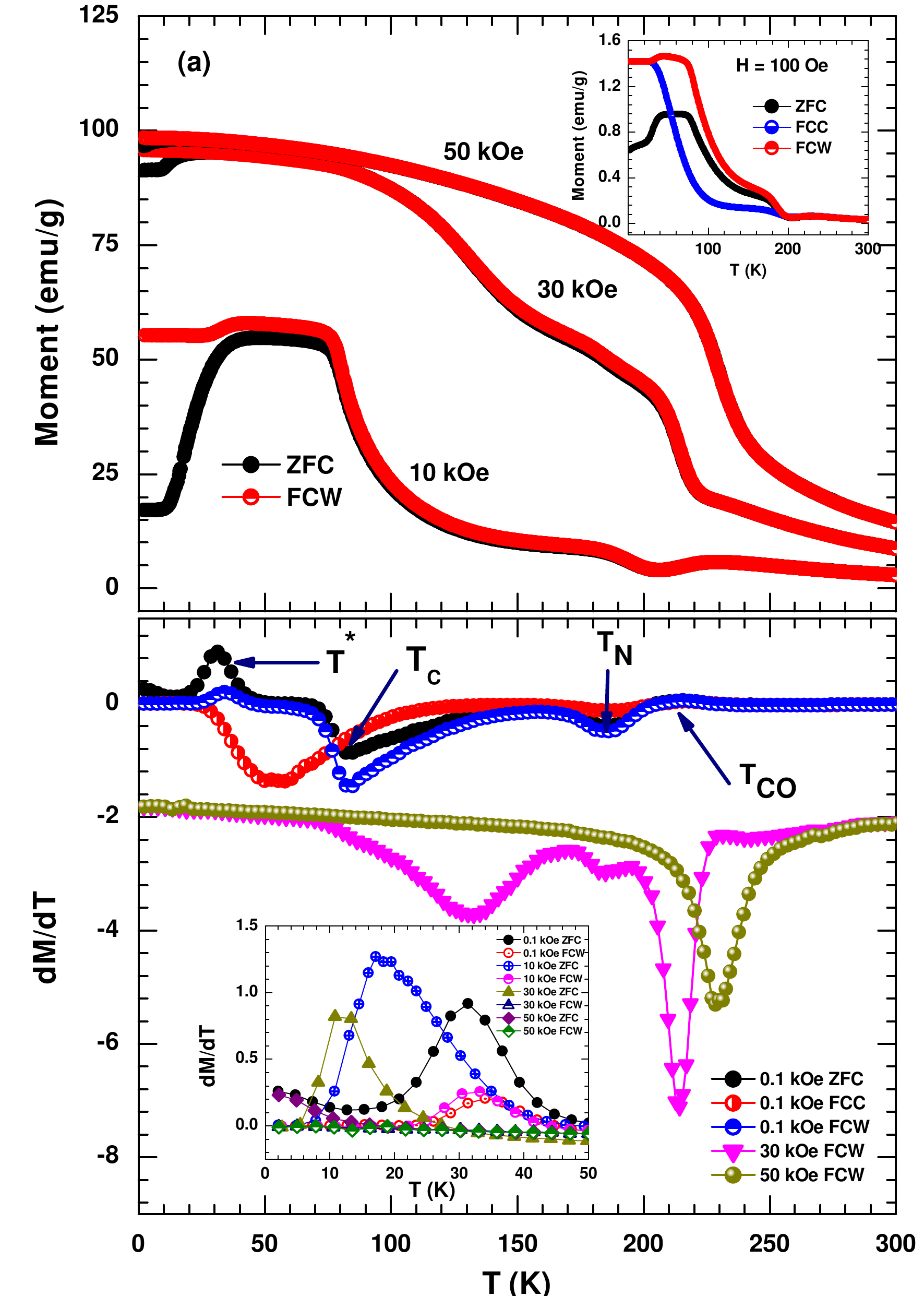}
 	\caption{(a) Temperature dependence of ZFC and FCW magnetization of La$_{0.2}$Pr$_{0.4}$Ca$_{0.4}$MnO$_{3}$ (LPCMO) sample at various applied magnetic fields. Inset show ZFC, FCC and FCW magnetizations for $H=100$\,Oe. (b) Derivative plots of magnetizations. Inset highlights the temperature/magnetic field dependence of transition at $T^{*}$.}
 	\label{fig:ZFC-FCC-FCW}
 \end{figure}

The inset of Fig.~\ref{fig:ZFC-FCC-FCW}(a) shows the zero-field-cooled (ZFC) and field-cooled (in cooling (FCC) and warming (FCW) modes) magnetization data for an applied magnetic field of 100\,Oe in the temperature range 2--300\,K. As the temperature is decreased, the ZFC, FCC and FCW magnetizations start separating below 220\,K and indicate multiple magnetic transitions. In order to see these transitions clearly, we have plotted the derivative of the magnetizations in the main panel of Fig.~\ref{fig:ZFC-FCC-FCW}(b). In the ZFC and the FCW magnetizations, the transitions at 215\,K, 185\,K and 84\,K correspond to the charge ordering ($T_{CO}$), the antiferromagnetic ordering ($T_{N}$) and the ferromagnetic ordering ($T_{C}$), respectively. There is a clear difference between the FCW and the FCC magnetization curves; the ferromagnetic transition appears at $T_{C}\approx84$\,K in the FCW magnetization whereas the same appears at $T_{C}\approx55$\,K in the FCC magnetization. The hysteresis between the FCW and FCC magnetizations observed in our sample resembles that reported earlier in a related compound, La$_{0.225}$Pr$_{0.4}$Ca$_{0.375}$MnO$_{3}$ \cite{Macia} and is a characteristic of the phase separation (PS). In order to further investigate the nature of these magnetic transitions, we have measured the temperature dependence of  the magnetization for various applied magnetic fields (10\,kOe, 30\,kOe and 50\,kOe) which are shown in the main panel of Fig.~\ref{fig:ZFC-FCC-FCW}(a). The magnetization behaviour for $H=10$\,kOe is almost the same as that for $H=100$\,Oe; we can still see the signatures of the charge ordering as well the AFM ordering. However, as the applied field is further increased, the observed behaviour in magnetization changes drastically; for 30\,kOe, the CO is interrupted by the sudden appearance of a FM component (more evident in the derivative plot, which is shown in the main panel of Fig.~\ref{fig:ZFC-FCC-FCW}(b)) at $T\approx215$\,K and the original FM transition shifts to a higher temperature ($T_{C}\approx133$\,K). Existence of the AFM transition is evident from the derivative plot at $T_{N}\approx185$\,K. At 50\,kOe, one can see the fingerprints of only one transition at $T_{C}\approx230$\,K. This is a clear indication of the melting of the CO state which leads to an increase in the FM ordering transition temperature as the applied magnetic field is increased. 
 
 \begin{figure}
 	\centering
 	\includegraphics[width=\linewidth, height=\linewidth]{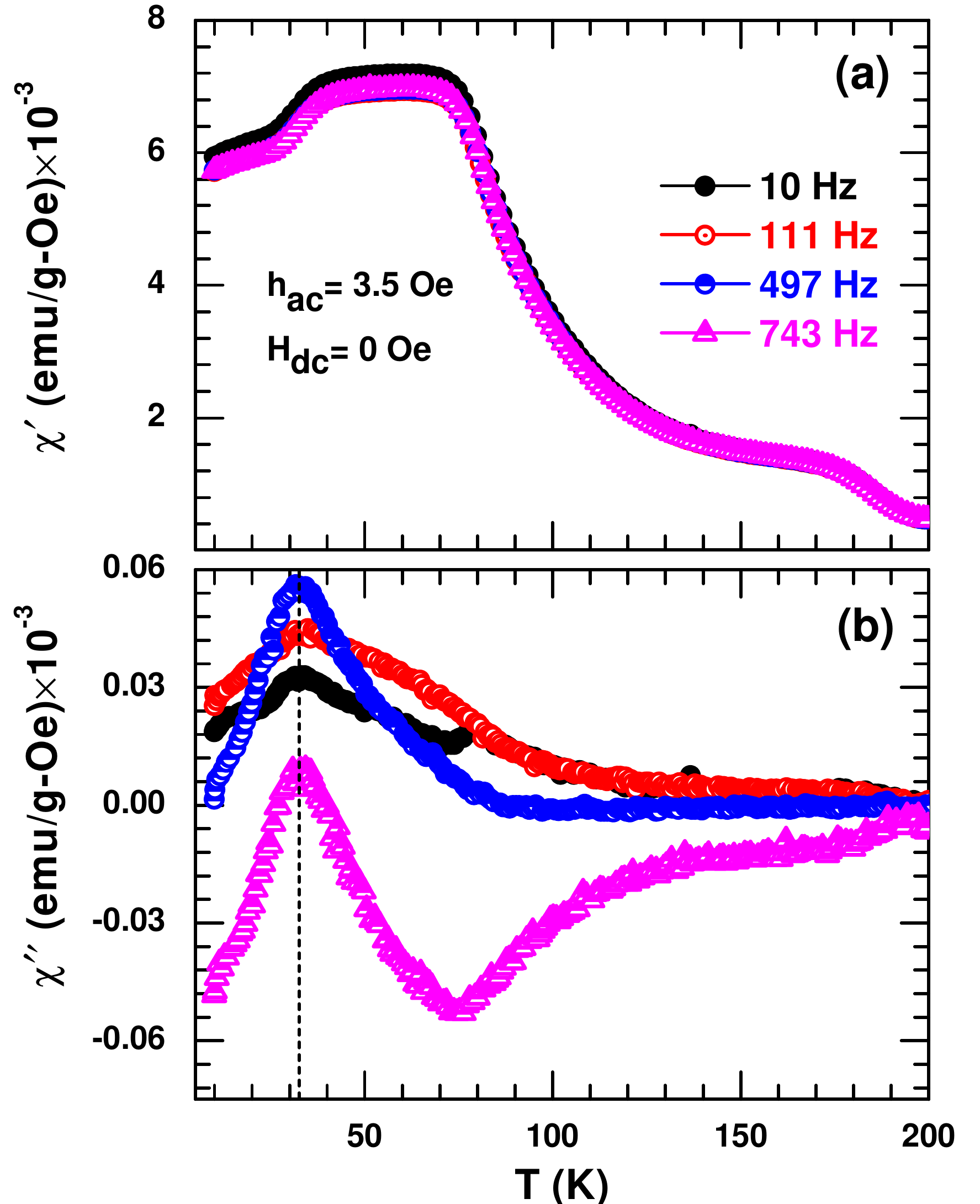}
 	\caption{Temperature dependence of ac susceptibility for La$_{0.2}$Pr$_{0.4}$Ca$_{0.4}$MnO$_{3}$ at different frequencies; (a) real part $\chi'(T)$ and (b) imaginary part $\chi''(T)$}
 	\label{fig:AC}
 \end{figure}

An additional magnetic transition is seen in the ZFC and the FCW magnetization data at low temperatures, marked as $ T^{*} $($\approx34$\,K for 100\,Oe) in  Fig.~\ref{fig:ZFC-FCC-FCW}(b), which is absent in the FCC data. A similar transition observed in another related compound, a single crystal of La$_{0.35}$Pr$_{0.275}$Ca$_{0.375}$MnO$_{3}$, was attributed to a spin-glass transition \cite{Srikanth} whereas this transition in a polycrystalline sample of La$_{0.225}$Pr$_{0.4}$Ca$_{0.375}$MnO$_{3}$ was explained on the basis of a blocking temperature at the phase separation \cite{Macia}. In order to investigate the nature of this transition, we have measured the ac susceptibility ($h_{ac}=3.5$\,Oe) of our sample as a function of temperature at different frequencies, as shown in Fig.\,\ref{fig:AC}(a) and (b) for in-phase $\chi'(T)$ and out-of-phase $\chi''(T)$ signals, respectively. We could not observe any differences between the ac susceptibilities measured at different frequencies especially corresponding to the glassy peak, which is further confirmed by the same peak position at $ T^{*} $ in the $\chi''(T)$ curves. Our results  indicate the absence of any spin-glass type transition in this compound. However, the magnetic transition at $T^{*} $ shows a clear dependence on applied magnetic fields. In the ZFC mode, this transition shifts to lower temperatures as the magnetic field is increased whereas in the FCW mode, it remains at the same temperature for 0\,kOe and 10\,kOe and disappears for higher magnetic fields (30\,kOe and 50\,kOe); (see the inset of Fig.~\ref{fig:ZFC-FCC-FCW}(b)).

\begin{figure}[hbtp]
	\centering
	\includegraphics[width=\linewidth]{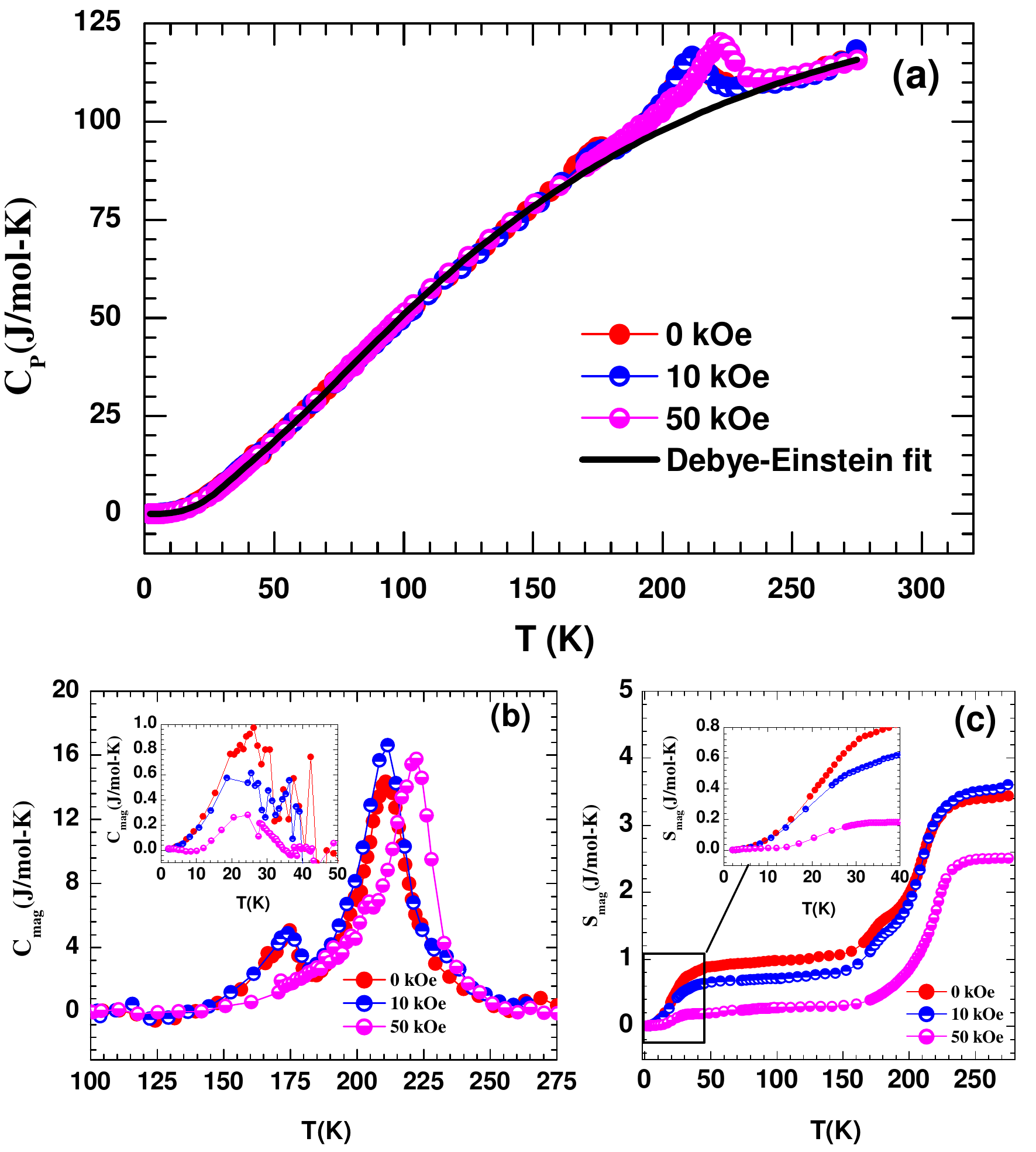}
	\caption{(a) Temperature dependence of total  heat capacity ($C_{P}$) of La$_{0.2}$Pr$_{0.4}$Ca$_{0.4}$MnO$_{3}$ (LPCMO) sample at different applied magnetic fields. Solid line represents the Debye-Einstein fit for phonon contribution. (b) Temperature dependence of magnetic contribution of heat capacity ($C_\text{mag}$) at various applied magnetic field. Inset show the $C_\text{mag}$ for 2--50\,K temperature range. (c) Variation of magnetic entropy ($S\mathrm{_{mag}} $) at different applied magnetic fields. The inset shows the expanded part of $S_\text{mag} $ at low temperatures.}
	\label{fig:HC}	
\end{figure}

The results obtained for the temperature variation of heat capacity at different applied magnetic fields (0\,kOe, 10\,kOe and 50\,kOe) are shown in Fig.~\ref{fig:HC}(a) in the temperature range 2--275\,K. In order to find the magnetic contribution in the total heat capacity ($C_P$), the phonon contribution  was estimated by fitting $C_P$ using the combined Debye and Einstein equations \cite{Koteswararao},

\begin{equation}
\begin{split}
C_{ph} = &\alpha_{D} \left[9k_{B}\left(\frac{T}{\theta_{D}}\right)^{3}\int_{0}^{x_{D}}\frac{x^{4}e^{x}}{(e^{x}-1)^{2}}dx\right]\\
&+\sum_{i}\alpha_{E_{i}}\left[3R\left(\frac{\theta_{E_{i}}}{T}\right)^{2}\frac{exp\left(\frac{\theta_{E_{i}}}{T}\right)}{\left[exp\left(\frac{\theta_{E_{i}}}{T}\right)-1\right]^{2}}\right]
\end{split}
\end{equation}

\noindent where $\alpha_{D}$ and $\alpha_{E_{i}}$ are the associated coefficients, $k_{B}$ is the Boltzmann constant, $\theta_{D}$ is the Debye temperature, $R$ is the universal gas constant, $\theta_{E_{i}}$ is the Einstein temperature and $x_{D}=\theta_{D}/T$. 
The best fit to the above equation, using one Debye term and three Einstein terms, is shown as solid line in Fig.~\ref{fig:HC}(a). The magnetic contribution to the heat capacity $C_{\text{mag}}$ was obtained by subtracting the phonon contribution from $C_P$ ($ C_{\text{mag}} = C_{P}-C_{ph} $), which is shown in Fig.~\ref{fig:HC}(b). For $H=0$\,kOe and 10\,kOe, two peaks are observed, one at $T\approx174$\,K and the other at $T\approx212$\,K. When the applied magnetic field is increased to 50\,kOe, we see only one peak at $T\approx222$\,K. Comparing with the magnetization data, we can assign (for $H=0$\,kOe and 10\,kOe) the peak at 212\,K as the peak arising due to the CO transition and the peak at 174\,K to the AFM ordering. We do not see a peak corresponding to the FM transition as expected for a smooth cross-over between the states (not a true phase transition). However, for $H=50$\,kOe, where we see only one peak in heat capacity, the peak at 222\,K can be assigned as the peak arising due to the FM transition as the CO would have melted. It is also possible that the CO and FM transitions occur at the same temperature and the peak appears due to the CO state as in the case for lower applied magnetic fields. 

In the magnetic contribution to heat capacity, we could observe a small hump at low temperatures (see inset of Fig.\,\ref{fig:HC}(b)), which gets suppressed slowly by the application of the magnetic field. This peak may be arising due to the $T^{*}$ transition that we observe in the magnetization data. The magnetic entropy $S_{\text{mag}} $ was evaluated by using the equation \cite{Tomy},

\begin{equation}
S_{\text{mag}}=\int_{T_{1}}^{T_{2}}\frac{C_{\text{mag}}}{T}dT
\end{equation}

\noindent The temperature dependence of the magnetic entropy of LPCMO sample is shown in Fig.~\ref{fig:HC}(c)) at different applied magnetic fields. The magnetic entropy increases with increasing temperature and saturates to a value of 3.3\,J/mol-K, 3.5\,J/mol-K and 2.45\,J/mol-K above 240\,K, corresponding to the applied magnetic fields of 0\,kOe, 10\,kOe and 50\,kOe, respectively. These observations are consistent with the magnetization data; the initial increase is due to the melting of the strain glass state into the strain liquid state leading to the plateau in the intermediate region (the FM transition occurs in this region without any peak in the heat capacity) before the magnetic/CO transitions start occurring.

\begin{figure}[hbtp]
	\centering
	\includegraphics[width=\linewidth]{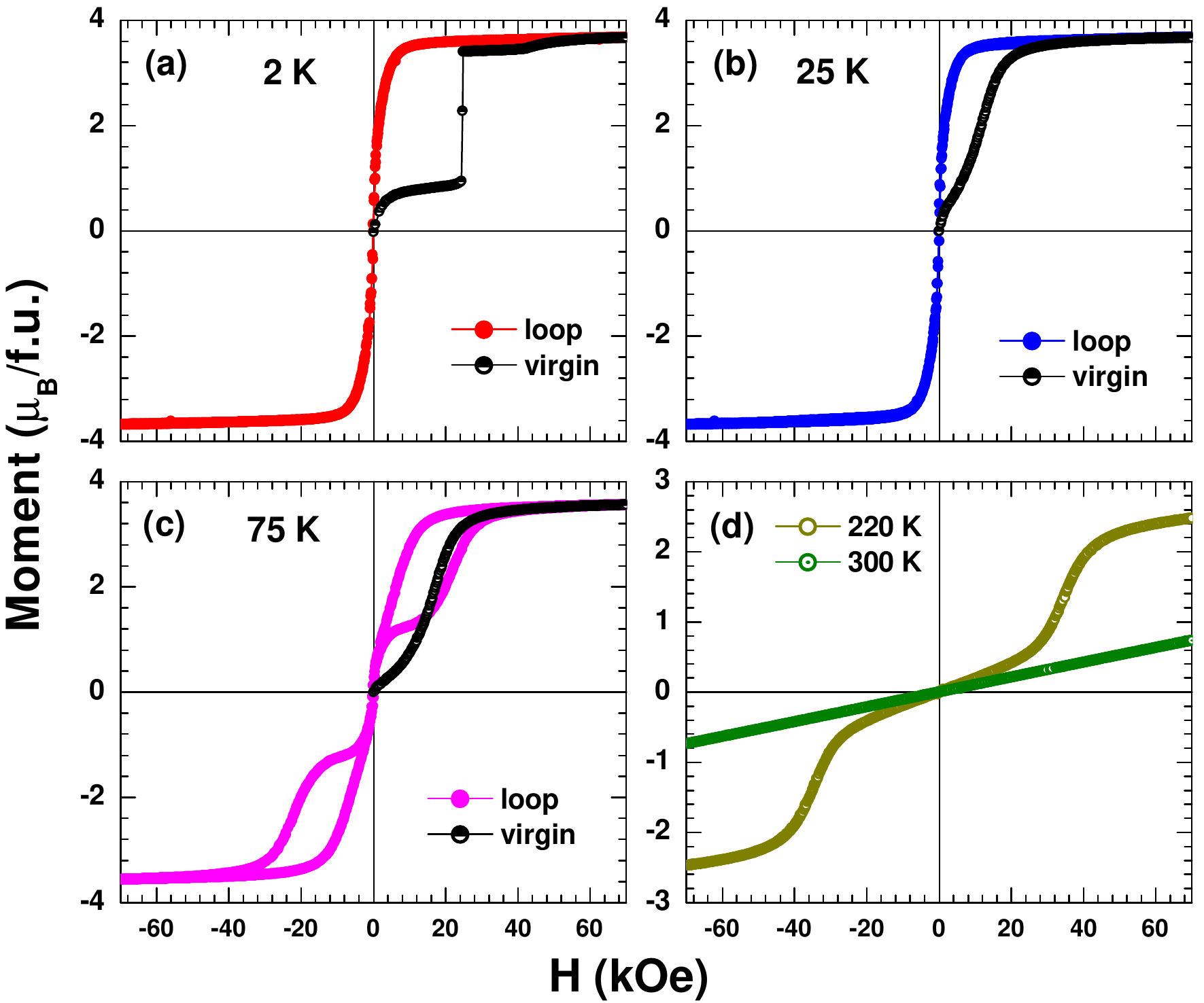}
	\caption{Magnetic field dependence of magnetization of La$_{0.2}$Pr$_{0.4}$Ca$_{0.4}$MnO$_{3}$ sample at selected temperatures (a) 2\,K, (b) 25\,K, (c) 75\,K and (d) 220\,K and 300\,K. The black solid symbols show virgin curve of the magnetization.}
	\label{fig:MH}
\end{figure}
 
We have measured the five quadrant $M$--$H$ curves for the LPCMO sample at different temperatures. After taking the $M$--$H$ data at each temperature, the sample was warmed above all the transition temperatures ($T \gg T_{CO}$) and then cooled again to the required temperatures in zero applied magnetic field. Figure~\ref{fig:MH} shows the magnetic hysteresis loops as a function of applied magnetic fields ($\pm$70\,kOe) at different temperatures. At $ T=2 $\,K, as the magnetic field is increased the virgin magnetization (black solid symbol) shows an ultra sharp metamagnetic transition from one FM state to another FM state at a critical field around $ H_{c}\approx 24.5$\,kOe and with the further increase in the magnetic field, one more metamagnetic transition appears at $ H_{c}\approx41.5$\,kOe. However, we do not observe any such metamagnetic transitions in other quadrants. Even though the hysteresis behaviour is almost similar upto 50\,K, the sharpness of the metamagnetic transition is diminished, the second metamagnetic transition is not distinguishable (we show a typical magnetization curve at 25\,K in Fig.~\ref{fig:MH}(b)) and the critical fields at which the metamagnetic transitions occur decreases, $ H_{c}\approx6$\,kOe at $T=25$\,K. As the temperature is further increased, the reversibility (excluding the first quadrant) in the $M$--$H$ curves starts disappearing as shown in Fig.\,\ref{fig:MH}(c) for $T=75$\,K. The metamagnetic transition appears in other quadrants also, in the 3$^{\text{rd}} $ and the 5$^{\text{th}}$, in addition to the first quadrant. We can also see a clear cross-over between the magnetization curves of the 1$ ^{\text{st}} $ and the 5$^{\text{th}}$ quadrants (see Fig.\,\ref{fig:MH}(c)). Even at $ T= 220$\,K, (slightly above all the transition temperatures), we can still see fingerprints of the metamagnetic transition, though the $M$--$H$ curves are completely reversible (in all quadrants) and at 300\,K, we observe only a linear response as expected for the sample in a paramagnetic state which is shown in Fig.~\ref{fig:MH}(d).
 
 \begin{figure}[hbtp]
 	\centering
 	\includegraphics[width=\linewidth]{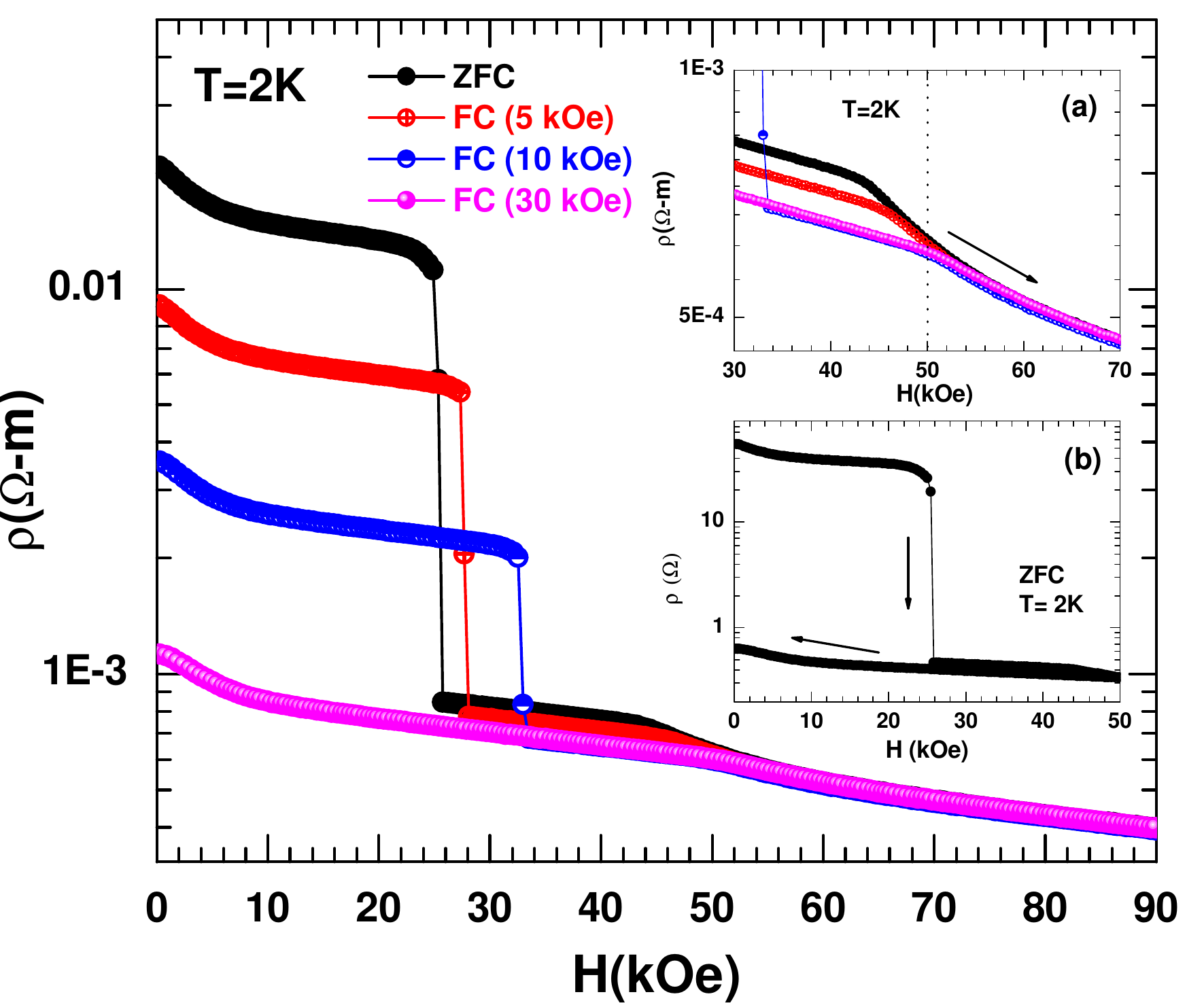}
 	\caption{Magnetic field dependence of resistivity at 2\,K in cooling at different applied magnetic fields. The inset (a) shows the closer view of the resistivity between 30\,kOe to 70\,kOe, measured at different applied fields and inset (b) shows the magnetic field dependence of resistance at 2\,K in ZFC mode.}
 	\label{fig:RH}
 \end{figure}
 
 In order to corroborate the sharp jump in magnetization arises from the conversion of COI phase into FMM phase, we have measured the resistance as a function of applied magnetic field at $T=2$\,K which is plotted in Fig.~\ref{fig:RH} in the ZFC and FC modes. In the ZFC mode, the resistivity data exhibits a sharp drop from a higher resistance state to lower resistance state at a critical field of 24.5\,kOe, which is consistent with the magnetization data where a sharp metamagnetic transition is observed. As the field is further increased, we see another smooth transition around 41.5\,kOe, which is also consistent with the second metamagnetic transition (see the inset (a) of Fig.~\ref{fig:RH}). If the samples are field-cooled, these sharp drops related to the first metamagnetic transition shift to higher field values (27\,kOe and 32\,kOe for FC(5\,kOe) and FC(10\,kOe), respectively) and for $H=30$\,kOe, we see only a shallow transition due to the second metamagnetic transition. The inset (b) in Fig.~\ref{fig:RH} shows the resistance in the increasing as well as decreasing magnetic field for a zero field-cooled sample. In the reverse cycle of the magnetic field change, we do not see the signature of any metamagnetic transitions, which is again in good agreement with the $M$–$H$ data. (Fig.~\ref{fig:MH}(a)). 
 
\section {Conclusion}
We have presented a comprehensive study in one of the phase separated manganites, La$_{0.2}$Pr$_{0.4}$Ca$_{0.4}$MnO$_{3}$, and elucidated some of the issues resulting from the phase separation in these compounds. The compound shows multiple magnetic transitions, which are magnetic field dependent; for an applied magnetic field of $ H= 50$\,kOe, the CO state is converted into a FM state with a large increase in $T_{C}$. The ac susceptibility measurements show that the transition at low temperatures ($ T^{*} $) does not depend on the frequency, thus indicating the absence of any spin glass behaviour. Magnetization as well as heat capacity measurements indicate that this low temperature transition is magnetic field dependent. In ZFC mode, the field dependent resistivity shows a sharp drop at 2\,K indicating that the sample behaviour changes from a high resistive state to a low resistive state, pointing towards the conversion of the COI phase to a FMM phase. Our results confirm the existence of phase separation, rigidity of the low temperature glassy-like phase as well as the conversion of COI phase to FMM phase by the application of magnetic fields.

\section*{Acknowledgment}
DK acknowledges Dr. Himanshu Sharma, Tohoku University Sendai, Japan for valuable discussion. DK would like to thanks the Council of Scientific and Industrial Research (CSIR), India for providing a senior research fellowship and also thanks to the Industrial Research and Consultancy Center (IRCC) IIT Bombay for providing the measurement facility.


\bibliographystyle{apsrev4-1}
\bibliography{bib}

\end{document}